# Phase-Retrieved Tomography enables imaging of a Tumor Spheroid in Mesoscopy Regime

Daniele Ancora[1,2,*], Diego Di Battista[1,2], Georgia Giasafaki[1], Stylianos E. Psycharakis[1], Evangelos Liapis[1] and Giannis Zacharakis[1]

[1]Institute of Electronic Structure and Laser, Foundation for Research and Technology Hellas, GR-70013 Heraklion, Greece.

[2]Department of Materials Science and Technology, University of Crete, GR-71003 Heraklion, Greece.

**ABSTRACT (150 words)**

Optical tomographic imaging of biological specimen bases its reliability on the combination of both accurate experimental measures and advanced computational techniques. In general, due to high scattering and absorption in most of the tissues, multi-view geometries are required to reduce diffuse halo and blurring in the reconstructions. Scanning processes are used to acquire the data but they inevitably introduces perturbation, negating the assumption of aligned measures. Here we propose an innovative, registration-free, imaging protocol implemented to image a human tumor spheroid at mesoscopic regime. The technique relies on the calculation of autocorrelation-sinogram and object's autocorrelation, finalizing the tomographic reconstruction via a three-dimensional Gerchberg-Saxton algorithm that retrieves the missing phase information. Our method is conceptually simple and focuses on single image acquisition, regardless of the specimen position in the camera plane. We demonstrate increased deep resolution abilities, not achievable with the current approaches, rendering the data alignment process obsolete.

**MAIN**

In recent years, almost every tomographic-based imaging process cannot abstract from the usage of computational techniques. Particularly in Biomedical Imaging at optical wavelengths, where light scattering plays an important role in terms of image quality, the link between experiments and algorithms is so tight that every improvement on one side rapidly affect the other. In fact, computational methods are nowadays becoming so important that they offer opportunities to develop better imaging methodologies or unlock possibilities to image in prohibitive conditions [ 1]. Although the tomographic principle is conceptually simple, i.e. optically slicing the sample for internal functional and structural inspection, it often requires big effort on both computational and experimental levels in order to achieve satisfying image quality. Among other features, optical techniques allow *in-vivo* three dimensional imaging of biological samples, ranging from measurements in the macroscopic regime with Fluorescent Molecular Tomography [ 2] down to the microscopic level with confocal and multiphoton microscopy [ 3 4]. In practice, Selective Plane Illumination Microscopy (SPIM) [ 5 6] is one of the most widely used techniques that allows direct and real time [ 7] optical slicing of the sample and visualization of its internal structures *in-vivo*. The illumination of the specimen is accomplished with a light sheet and different camera images are collected, scanning the sample and stacking one slice next to the other. In such a way the tomographic volume is built with very little computational effort, in particular in cases of optically transparent or cleared [ 8] specimens. Although the technique works reasonably well in the microscopic regime, challenges arise in mesoscopy where higher scattering and absorption impede uniform and localized illumination or emission, resulting in shadows and blurring at the camera detection level. Many approaches were proposed to tackle these problematics, ranging from combining multiple projections at different angles [ 5] to pivoting the light sheet with double illumination [ 9], using multiple multi-view geometry [ 10], or mixed approaches based on forward light modelling such as Mesoscopic Fluorescence Tomography (MFT) [ 11]. All of them require dedicated care for alignment and co-registration processes during both experimental and post-processing stages. Even the very robust and easy to implement Optical Projection Tomography (OPT) technique [12], based on the acquisition of projections at different angles, needs to take into account possible misalignments of the measuring scheme. Among other challenges and problems [13 14 15], the possibility that the sample could potentially exit the field of view along with translational and rotational misalignment drastically deteriorate the quality of the inverse reconstruction. Computational techniques for finding the Center of Rotation (CoR) [16], correcting for sample movements [17 18] and selecting an appropriate Region of Interest (RoI) [19], are often used in order to overcome these problems, although they share the disadvantage of being complicated and sensitive to each specific measurement.

When scattering becomes relevant, at around one Transport Mean Free Path (TMFP) in the so-called mesoscopic regime of imaging [1], a tool for universal data reinterpretation is still required to study and characterize model specimens of high scientific interest, such as *tumor spheroids*, *Drosophilae melanogaster* (at larval stages) and *Zebrafish*.

In this work we focus on three-dimensional image reconstruction of early stage necrosis distribution in a human-breast tumor spheroid in the mesoscopic regime of light scattering. By combining the functional information from SPIM and structural information from OPT in a complementary fashion, we propose a novel "escamotage" to uniquely align the dataset by exploiting the mathematical properties of image autocorrelation. Interestingly, autocorrelation-based imaging is currently being employed in novel promising applications for allowing, under certain conditions dictated by the memory effect range [20], visualization through turbid media and behind corners [21 22].

In biology, the multi-cellular tumor spheroid (MCTS) is the best characterized and most widely used scaffold-free 3D culture system that takes advantage of the inherent ability of many cancer cells to self-organize into spherical clusters [23 24 25]. MCTS is gaining huge attention as a pre-clinical drug-testing model and a large body of literature over the past three decades highlights the usefulness of this model system in translational cancer research and drug discovery. Herein, we propose an innovative tomographic approach, based on the reconstruction of the three dimensional sample's autocorrelation rather than on direct imaging of the specimen itself. The fluorescence emitted by the spheroid necrotic cells is successfully reconstructed by solving the phase retrieval problem related to its three-dimensional autocorrelation, without any need for data alignment. Because of the novel usage of the algorithm combined with SPIM-OPT approach, we found appropriate referring to this approach with the name Phase-Retrieved Tomography (PRT).

Employing PRT, we uniquely imaged the fluorescence distribution of the cell-death marker DRAQ7$^{TM}$ in a T47D human breast tumor spheroid with a diameter of about 200µm. A schematic of the experimental measurements performed in a combined SPIM-OPT setup is illustrated in Fig. 1: per each rotation angle, we acquired a collection of tomographic slices by illuminating the sample with a light sheet perpendicular to the camera plane. We placed the specimen inside a Fluorinated Ethylene Propylene (FEP) tube and we measured it over a complete rotation, making sure the whole sample was illuminated during each SPIM acquisition. While rotating, in case the spheroid exited the field of view, we translated the object along the detection focal plane (x-axis) back to the center of the camera, without being concerned for such a displacement. For each SPIM dataset, at every angle of rotation, we calculated the Maximum Intensity Projection (MIP), stacking them one after the other as it would have happened in an OPT experiment (MIP-sinogram). The measurements were performed in the mesoscopic regime of light scattering, as can be observed in Fig. 2. In fact the sample is non-uniformly absorbing at visible wavelengths (bright field image in Fig. 2A) and it is big enough to scatter the fluorescent light excited on the side opposite to the camera detection line. In fact, in Fig. 2 (panels B-E) it is notable that groups of fluorophores are highly blurred at certain angles, while these become evident after a rotation of 180°. Therefore, in these circumstances, it is crucial to exploit all the information coming from a full rotation of the sample to accurately retrieve even hidden fluorophore distributions. Additionally in this regime of scattering and for such a small specimen, a classical sinogram-based reconstruction approach is strongly sensitive to mechanical vibrations and off-axis rotations. The sample drift is clearly visible when comparing the projections at 0° and 360° (bottom of Fig. 3A) and yields to a misaligned MIP-sinogram (Fig. 3B). With our stage the total diagonal drift after a complete rotation was 18.7 µm, 17.6 µm and 6.4 µm along x and y direction in the camera image plane, which is about 10% of the whole diameter of the spheroid. In this difficult scenario, using simple camera projections (MIP) will turn into a misaligned sinogram which would have to be carefully post-processed in order to correct for such displacement. In our experiment instead, we calculated the 2D autocorrelation of each MIP, stacking them in the same order and assuming known the rotating step. The result of this calculation is a new autocorrelation sinogram (A-sinogram) which is always perfectly centered, regardless of where the spheroid was in the camera plane (Fig. 3D). The simple backprojection of the A-sinogram, via inverse Radon transform (Supplementary Materials), leads to the calculation of a volumetric dataset corresponding to the three-dimensional autocorrelation of the sample shown in Fig. 4A. We proved numerically this fact by calculating the 3D autocorrelation of a commonly used test-object, a three-dimensional Shepp-Logan phantom (Supplementary Materials), and comparing it with what found after the backprojection of its A-sinogram. Then retrieving the phase information from such a volume, i.e. the three-dimensional autocorrelation of the real object, is a typical phase retrieval problem with higher dimensionality [26], here for the first time employed for tomographic purposes. The complexity of the algorithm poses computational challenges, which are substantially related to three-dimensional Fourier Transformations and can be easily overcome by parallel GPU implementation. In our specific case a normal GPU nVidia, 2-years old GeForce 780Ti with 2880 CUDA cores, could handle a volume up to $300^3$ voxels in direct space before running out of memory, a technical limitation that can be easily fixed in future algorithm developments. The result of the PRT imaging reconstruction is the volume shown in Fig. 4B, which is the indirect reconstruction of the DRAQ7™ fluorescence distribution within the tumor spheroid. To validate the PRT reconstruction we compared it with classical OPT reconstruction and multi angle SPIM measurements. Figure 5D-F shows quantitative results for the PRT technique by visualizing the color-coded MIP (hyperstack, in which the color indicate the depth) of the whole dataset (Fig. 4B) seen from different angles. The classical Radon transform of the sinogram of Fig. 3B, co-registered by finding its CoR, leads to not satisfying reconstructions (Fig. 5A-C) in which we cannot distinguish any single necrosis, thus making classical approaches not usable for this imaging regime. Neither direct SPIM imaging (Fig. 2B-E) gave any better performances, resulting in blurred region for some groups of fluorophores found deep inside in the spheroid (underlined by dashed circles). PRT imaging instead, retrieved a highly detailed fluorescent distribution (Fig. 5D-E), visualizing every single cell even those that were found blurred in SPIM measurements at various projection angles (dashed circles). The correct coloring sequence of the dataset, hypeuniform stack visualization using the Fiji toolbox [27], efficiently displays correct distances between fluorophores compared to those of SPIM and OPT. This is a natural consequence of the correct data backprojection in the autocorrelation domain, which does not need any alignment process. The phase retrieval algorithm always returned comparable reconstructions, regardless of the initial and random starting guess for the phase, thus making the PRT results more efficient than the other currently used techniques. Interestingly the PRT technique could also be used for the restoration of previously acquired dataset, which did not lead to any proper imaging due to eventual misalignment of the system. Because of these evidences, we believe that PRT can enter convincingly in the current biomedical imaging scenario, in particular as an efficient tool for correct data reinterpretation in highly sensitive measurements.

**DISCUSSION**

Throughout the whole text, for the first time in biomedical imaging, we presented the possibility to adopt a three-dimensional autocorrelation reconstruction in combination with a phase retrieval algorithm to outperform three-dimensional tomographic imaging of a biological specimen. A human tumor spheroid was successfully imaged in the mesoscopic regime by using a combined SPIM-OPT setup to reconstruct the object by using the phase retrieval algorithm in a novel fashion. We refer to this innovative approach as Phase-Retrieved Tomography (PRT) because we are retrieving the phase related to the autocorrelation, and so with the Fourier modulus, that of the whole object. The new approach is absolutely insensitive to the specimen translational misalignment and to stage drifts in the three spatial directions, only requiring prior knowledge of the rotational degree to correctly backproject the autocorrelations sinogram. We proved this numerically with a 3D Shepp-Logan phantom, showing its reconstruction to be quantitatively equal to the original sample and robust to vibrations of the acquired projection, even when perturbing the sinogram with random shifts in two directions in the camera plane (Supplementary Materials). Although the position of the retrieved object within the reconstruction volume is random, because of phase retrieval ambiguities [26], its signal distribution is always consistent with the real object and which implies reconstructions not affected by data collection misalignments. The experiments confirmed the same behavior and exceptional results were achieved by collecting MIPs of full rotation in steps of 2°. Consistent reconstructions can also be achieved with bright field illumination in a classical OPT approach, as the spheroid depicted in Fig. 1 was in fact retrieved with PRT in rear bright field illumination, demonstrating the flexibility of the method with different acquisition techniques.

Of more interest are the results obtained by PRT, which fully exploit the SPIM-OPT setup. At the moment in fact, one of the disadvantages of the SPIM technique is its poor resolution along the perpendicular direction with respect to the planar illumination (axial resolution). This can be improved with light sheet deconvolution [28 29] or co-registration of stacks at different angles [

30], both of which enhance the resolution along the third directions of the reconstruction, but in general share the disadvantage of being highly specific for each set of measures and sensitive to drifts. Even in the OPT field of study, the hurdles of recovery of the Center of Rotation (CoR) of the specimen is still an unsolved question. Although many approaches were proposed for data post-alignment, these fail to provide a generalized method for removing artifacts due to misalignments. Our method overcomes all these issues by retrieving the object from several SPIM-MIP autocorrelations rather than direct projections, which allows the obtained autocorrelation sinogram to be inherently aligned (Fig. 3 panel D) and rotate around its center of symmetry, regardless of the original object's rotating axis position. With our method there is no need for estimate CoR or RoI from the sinograms, leaving the user free to focus on pure, single measurements instead of aligning the rotational system and the dataset acquired. It is worth to notice that the autocorrelation of the projection at 0° always exactly matches the one at 360° even if the sample, while rotating, did not match the original starting position. The A-sinogram is used to compute the object three-dimensional autocorrelation, borrowing this ability from the classical OPT approach. In such a way, it is possible to also overcome the OPT weakness, i.e. the error in tracking the object while rotating, fulfilling, thus, the inverse Radon transform requirements and leading to the calculation of a nearly exact three-dimensional autocorrelation. Moreover, other iterative inversion techniques, such as ART, SIRT, and SART [ 31] can be straightforwardly implemented for the autocorrelation reconstruction, offering further room for improvement and applications for PRT.

Quantitative results are presented in Fig. 5 where the results are compared with the classical OPT reconstruction and where the coloring code represents the relative depth with respect to the first fluorescent signal. The artifacts due to misalignment of the data (Fig. 5A-C) disappear in the PRT reconstruction, which convincingly demonstrates its substantial tomographic capabilities, obviating to the need for data postalignment (Fig. 5D-F). SPIM hyperstacks at four perpendicular angles in Fig. 3 can also be compared to PRT reconstructions viewed from the same angles (x and y projections in the volume coordinates). The correct coloring sequence for each fluorophore validates the results of the PRT, exhibiting an enhanced deep resolution compared to that of SPIM. In fact, an interesting feature arises from the SPIM-MIPs underlined by dashed circles: in the mesoscopic regime, SPIM cannot resolve at every degree the presence of some group of fluorophores located on the opposite face of the spheroid, resulting in blurred regions in the image. PRT reconstruction, instead, clearly retrieves these objects (dash circles in Fig. 5) by fully exploiting the information coming from multiple angle MIP and exactly combining them in autocorrelation space for enhanced deep resolution ability. The tomographic projection of the PRT data shows comparable resolution capabilities with respect to the other directions visible in Fig. 3E. As a result, the novel method proposed here is quantitative, robust and easy to implement in a regular SPIM-OPT setup. Furthermore, it can be implemented in a normal OPT approach using every kind of illumination regime (bright field, fluorescence) and can also be extended to other projection based approaches, such as X-CT or SPECT. Technically, in those cases the spheroid, i.e. the whole object, would have to lie within the camera's depth of field, otherwise defocused images would be used to calculate the autocorrelations. In our scanning approach, the plane illuminated is always in focus and its MIP is blurred only because of internal scattering, allowing, thus, better reconstructions.

Finally, our method can accept further improvements in terms of quality of the Phase Retrieval algorithms used, which could potentially enhance even more its accuracy and its convergence rate, pushing the community towards developing faster three-dimensional implementations. In this scenario, it is worth taking into account another key aspect of this work. Because of its design, the PRT protocol can potentially be implemented in for imaging hidden three-dimensional specimens behind scattering curtains or around corners. Optical Imaging in such extreme conditions has been performed only with two-dimensional objects [ 21 22] by exploiting the speckle's autocorrelation property which, within the memory effect [ 20] regime, turns out to be identical to the autocorrelation of the object itself. The conceptual idea behind these studies is based on the fact that the autocorrelation of the signal is preserved while being scrambled by the scattering media. The so produced speckle pattern retains the autocorrelation properties of the object and it is related to its Fourier Transform (FT) modulus. The autocorrelation of the speckle pattern produced in front of the turbid layer can feed a Gerchberg–Saxton algorithm [32] used to retrieve its Fourier phase, allowing the reconstruction of the hidden object. Although mathematically the phase retrieval process works particularly well at every dimensionality [26], except for the lack of uniqueness in 1D problems, for optical imaging purposes (to the best of our knowledge) it has been used only in 2D implementations. In principle, following these considerations, the PRT method could tackle the current lack of hidden 3D imaging techniques, by accepting speckle patterns acquired at different angles rather than single projections, retaining its ability to perform valid reconstructions of the object embedded in an opaque curtain with no difference in the methodology proposed. With this work, we believe that a new path has opened toward biomedical imaging in diffusive media by virtue of the use of autocorrelation for tomographic purposes. Of course, further experimental prospects and novel questions emerge from this study, such as the effectiveness in perform hidden 3D imaging and a solid theoretical support for the method. We aim to address both by following studies in the near future together with other related questions, in order to widespread the applicability of phase-retrieved based tomography throughout, and not only, the biomedical imaging community.

**METHODS**

**SPIM-OPT measurements.** The experimental work presented was entirely accomplished using a combined SPIM-OPT setup, in which the sample position can be software controlled along the three spatial directions and rotated along an axis perpendicular to the camera detection plane by motorized stages. The sample was imaged acquiring 180 MIP projections (2° angle step) in order to complete a full rotation. At each projection the sample was scanned through the light sheet in steps of 20 µm while recording with the camera. The light sheet used has full width half maximum of 12 µm and the scanning speed was 20 µm/s.

For illumination, a continuous wave 635 nm diode laser was used. The light sheet was shaped by cylindrical optics and then was introduced vertically to the detection axis, having its central plane inside the focal plane of the 10x/0.28 infinity corrected detection objective (Mitutoyo, Japan). Finally for image acquisition a tube lens and an electron multiplying CCD (Ixon DV885, Andor Technology, Belfast,UK) were used. The resulting pixelsize of the system was 0.8 µm.

**Tumor spheroid generation.** Spheroids were generated with the hanging drop method using the Perfecta3D 96-well hanging drop plates (3d Biomatrix, Ann Arbor, MI, USA) following manufacturer's instructions. Briefly, cell suspensions for hanging drop experiments were made by dissociating cells with 0.5% trypsin-EDTA (Gibco, Grovemont Cir, Gaithersburg, MD, USA). Cell density was estimated using a hemocytometer. Dissociated cells were centrifuged at 1200 rpm for 5 min at room temperature, re-suspended in growth medium and diluted to a final concentration of 12.5 cells/µl. A 50 µl cell suspension was dispensed into each well of the spheroid culture plate to achieve an initial seeding density of 625 cells/well. In order to prevent evaporation, 1% agarose was added to the peripheral reservoirs of the hanging drop plates. The growth media was exchanged every other day by removing 25 µl of solution from a drop and replacing with 25 µl fresh media into the drop.

**Spheroid preparation for SPIM imaging.** 4 day old T47D breast tumor spheroids were incubated at 37 °C with 1.5 µM DRAQ7™ (Biostatus, Leicestershire, UK) for 24 h prior to imaging. DRAQ7™ is a far-red membrane impermeable fluorescent DNA dye that selectively stains the nuclei in dead and permeabilized cells. Staining of the spheroid with the nuclear dye was performed on the hanging droplet by replacing 10 µl from each hanging droplet with 10 µl 5x concentrated solution of the dye. Following, spheroids were transferred from the hanging drop plate to a microscope slide, washed twice with PBS and reconstituted in 100 µl Cygel Sustain (Biostatus, Leicestershire, UK) inside a cold room (4° C) to avoid rapid solidification of the Cygel. Then, the Cygel-embedded spheroid was transferred into a FEP tube (800µm inner diameter, Bola, Germany) which was sealed with blu-tack loaded on the SPIM instrument. The FEP tube containing the immobilized spheroid was embedded into a 37 °C water bath throughout the duration of the experiment to avoid liquefaction of the Cygel. The live spheroid was imaged in our custom SPIM setup using a diode laser for excitation (635 nm). The emission wavelength of the DRAQ7™ fluorophore peaks at 685 nm and, accordingly a 650 nm long pass filter was used to detect the fluorescence signal.

**3D Autocorrelation.** For each SPIM dataset at different angles, we calculated the Maximum Intensity Projection (MIP) of every frame. The images were cropped with a squared window of 300 pixels (field of view of 240 µm) containing the whole spheroid fluorescence signal. This results in a trembling sinogram, impossible to backproject with standard approaches (Fig. 3C). For each of the cropped MIP we calculated its autocorrelation with the Wiener-Khinchin theorem, stacking all of them in the same order. In this way we obtain the autocorrelation sinogram (A-sinogram) which is always aligned. This data was backprojected with the inverse Radon (using a Ram-Lak filter) transform function in MATLAB, obtaining a cubic volume with side of 599 pixels. This volume is the three-dimensional autocorrelation of the specimen of interest. No further post processing of the data was performed.

**Phase Retrieved Tomography.** The three-dimensional autocorrelation was used as a starting point for the Phase Retrieval problem. The reconstructing window within the autocorrelation volume had the size of the object. A mixed Hybrid Input-Output approach was used for 5000 steps followed by 1000 steps of Error-Reduction. The program was implemented in MATLAB with GPU-CUDA extension and typical running time for the reconstruction was about one hour.

**ACKNOWLEDGEMENTS**

**Funding.**

The GSRT Aristeia project "Skin-DOCTor" (1778); "Operational Programme Education, the Lifelong Learning" project "Neureka!", "Supporting Postdoctoral Researchers", co-funded by the European Social Fund (ESF) and National Resources. The EU Marie Curie ITN "OILTEBIA" PITN-GA-2012-317526, the FP7-REGPOT CCQCN (EC-GA 316165), the H2020 Laserlab Europe (EC-GA 654148) and the QNRF project No. NPRP9-383-1-083.

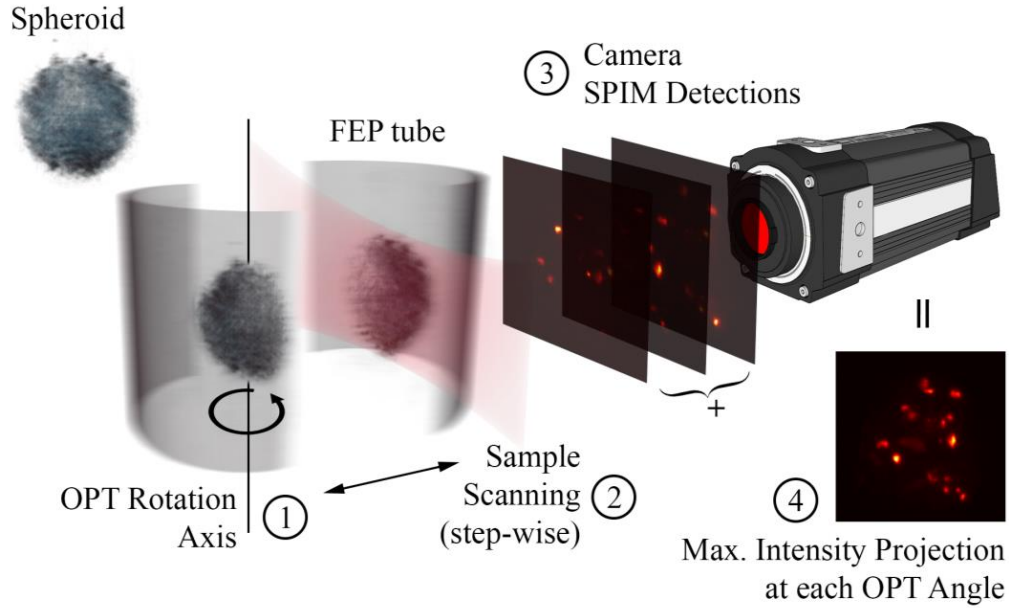

**Figure 1. | Schematic depiction of the data acquisition for the SPIM-OPT setup.** Starting from the left a tumor spheroid, stained with the DRAQ7™ fluorescent dye, is inserted into a FEP tube and mounted to the rotation stage. (1) The specimen is rotated at a known angle and then (2) is scanned through the light sheet along the detection axis. The illumination light sheet is established orthogonally to the detection axis and it fulfills the objective focal plane, in order to reduce out of focus contribution during the camera acquisition. While translating the specimen a SPIM detection (2D images stack) is stored (3). Finally (4), the data acquired is used to calculate the Maximum Intensity Projection (MIP) as a function of the angle. This procedure is repeated by rotating the sample (1) in steps of 2° until performing a complete rotation.

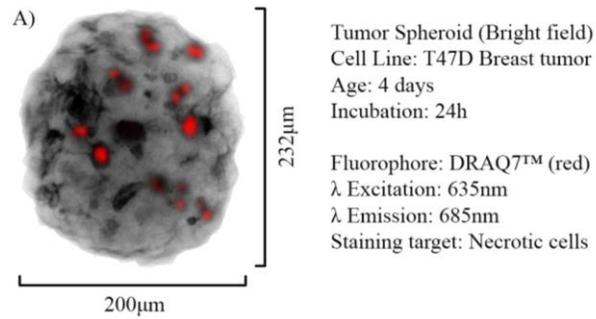
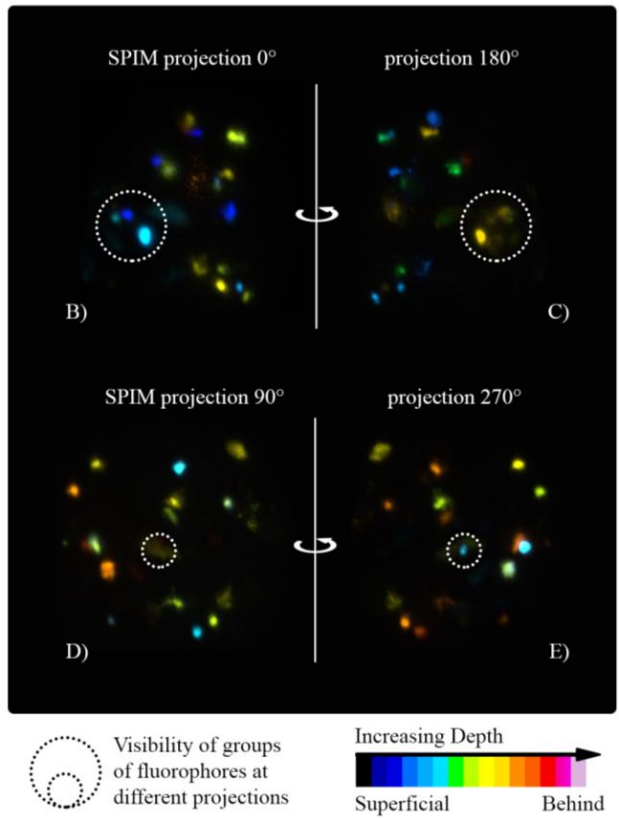

**Figure 2. | Imaging the spheroid in Mesoscopic regime. A**, Bright field image of the spheroid imaged at 0°. It is noticeable the non-uniform structure of the tumor mass. **B-E**, SPIM-MIP at perpendicular angles with respect to one another. Dashed circles point out groups of fluorophores not visible at some angles of measurements, due to the mesoscopic regime at which the spheroid was imaged.

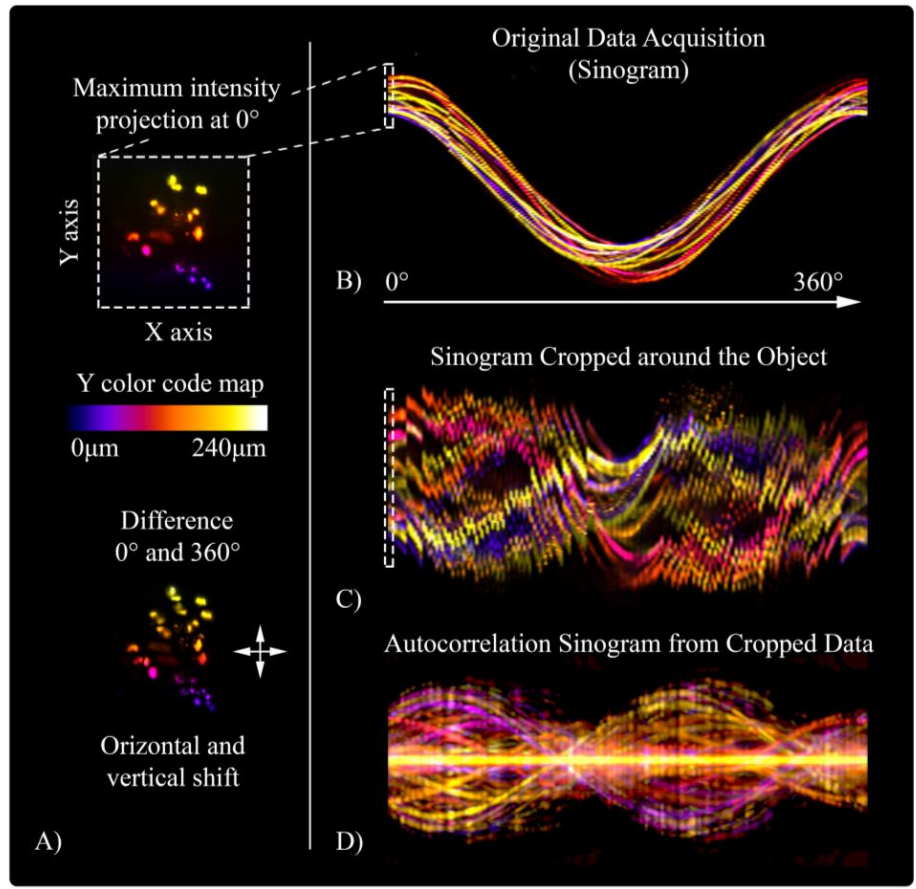

**Figure 3. | Sinogram based analysis of the acquired dataset. A**, MIP at 0° and relative difference after a full 360° rotation. The bidirectional drift of the sample during the measurements is visible on the bottom of panel **A**, and imply a misaligned MIP-sinogram. **B**, Original sinogram of the SPIM-MIP measurements as a function of the angle, the color code that labels the Y-axis depth is the same as in panel **A**. It is worth to notice that the color of the sinogram turns toward blue-red at the end of the rotation, which mean that the spheroid is slightly moving towards the bottom of the FEP tube. **C**, Cropped sinogram around the spheroid used to reduce the size of the reconstructed volume. It is worth noticing how the data is completely misaligned. **D**, Aligned autocorrelation sinogram calculated by using the data from the sinogram in **C**. The alignment of the dataset is achieved by simply calculating the autocorrelation for each MIP projection and stacking of the projections one after the other.

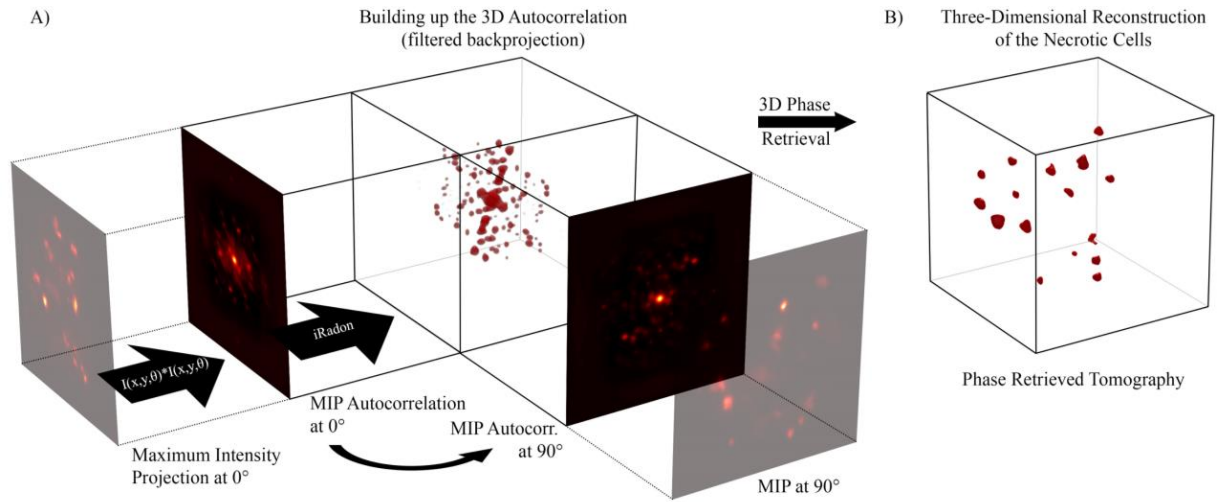

**Figure 4. | Schematics of the PRT approach. A**, schematic showing the backprojection criteria. For each of the Maximum Intensity Projections coming from every angle (Fig. 1) we calculate the autocorrelation images. These images are smeared along a volume in function of their angle of view, following the backprojection criteria of the filtered Inverse Radon transformation. The result of this is a volume that contains the three dimensional autocorrelation information of the object we want to image. It worth to notice that the autocorrelations are always peaked in the center and symmetric respect this point. In panel **B**, the final reconstruction after the Phase-Retrieved Tomography (PRT). The 3D autocorrelation feed a Gerchberg-Saxton algorithm, which retrieves the phase information reconstructing the object with no artifacts due to misalignment.

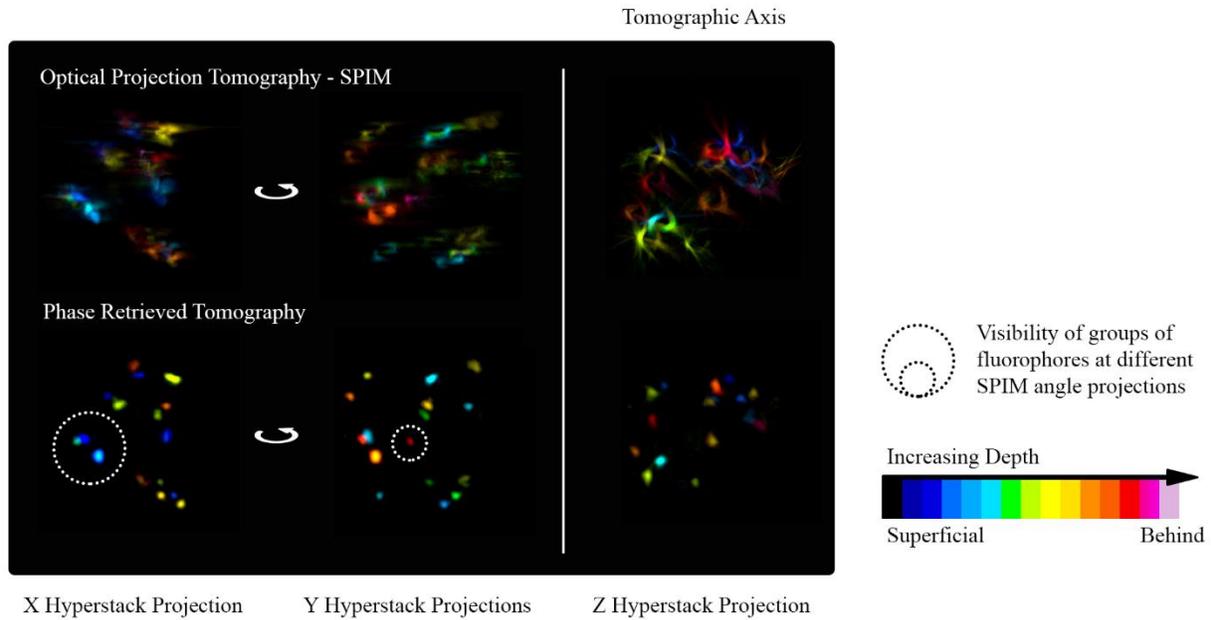

**Figure 5. | Imaging performance of the PRT technique**. Hyperstack projections for the SPIM-OPT measurements (panels **A**, **B**, **C**) and the PRT reconstructions (panels **D**, **E**, **F**). The color code denotes the depth at which the fluorophore is located, starting from the first signal. It is important to notice that the two methods share the same coloring order and the same fluorophore distribution, validating our reconstruction. Moreover, PRT shows a fluorophore while SPIM cannot resolve it, due to its location on the other side of the spheroid (red to white in the color scale). The Z hyperstack suggests comparable resolutions between the two methods along the tomographic axis.